# Understanding links between water-quality variables and nitrate concentration in freshwater streams using high-frequency sensor data

KERMORVANT Claire[1], LIQUET Benoit[1,2], LITT Guy[3], MENGERSEN Kerrie[4,5], PETERSON Erin E.[4,6], HYNDMAN Rob J.[7], JONES Jr. Jeremy B.[8], and LEIGH Catherine[4,9]

[1] CNRS/Univ Pau & Pays Adour, Laboratoire de Mathématiques et de leurs Applications de Pau - Fédération MIRA, UMR 5142, E2S-UPPA, 64600 Anglet, France
[2] Department of Mathematics and Statistics, Macquarie University, Australia
[3] Battelle, National Ecological Observatory Network, Boulder, CO 80301, United States
[4] ARC Centre of Excellence for Mathematics and Statistical Frontiers, Australia
[5] School of Mathematical Sciences, Queensland University of Technology, Brisbane, Australia
[6] Peterson Consulting, Brisbane Australia
[7] Monash University, Clayton, Victoria 3800, Australia
[8] Institute of Arctic Biology, and Department of Biology and Wildlife, University of Alaska Fairbanks, Fairbanks, AK 99709
[9] Biosciences and Food Technology Discipline, School of Science, RMIT University, Bundoora, Victoria 3083, Australia

**Key Points:**

- Big data from sensors in streams were analysed to understand links between nitrate concentration and other water-quality variables.
- An adequate set of water-quality variables were consistently important for understanding nitrate variation in very different sites.
- When the goal is to better understand nitrate dynamics, managers may base their monitoring programs on these variables.

Corresponding author: KERMORVANT Claire, `claire.kermorvant@univ-pau.fr`

**Abstract**

Real-time monitoring using *in-situ* sensors is becoming a common approach for measuring water-quality within watersheds. High-frequency measurements produce big data sets that present opportunities to conduct new analyses for improved understanding of water-quality dynamics and more effective management of rivers and streams. Of primary importance is enhancing knowledge of the relationships between nitrate, one of the most reactive forms of inorganic nitrogen in the aquatic environment, and other water-quality variables. We analysed high-frequency water-quality data from *in-situ* sensors deployed in three sites from different watersheds and climate zones within the National Ecological Observatory Network, USA. We used generalised additive mixed models to explain the nonlinear relationships at each site between nitrate concentration and conductivity, turbidity, dissolved oxygen, water temperature, and elevation. Temporal auto-correlation was modelled with an auto-regressive–moving-average (ARMA) model and we examined the relative importance of the explanatory variables. Total deviance explained by the models was high for all sites (almost 99%). Although variable importance and the smooth regression parameters differed among sites, the models explaining the most variation in nitrate contained the same explanatory variables. This study demonstrates that building a model for nitrate using the same set of explanatory water-quality variables is achievable, even for sites with vastly different environmental and climatic characteristics. Applying such models will assist managers to select cost-effective water-quality variables to monitor when the goals are to gain a spatially and temporally in-depth understanding of nitrate dynamics and adapt management plans accordingly.

# 1 Introduction

Nitrate is one of the most reactive forms of inorganic nitrogen in the aquatic environment (Camargo et al., 2005) and an essential component of the nitrogen cycle supporting life on Earth. In rivers, sources of nitrate include atmospheric deposition, groundwater, surface runoff and the biological degradation of organic matter present in freshwater ecosystems. In addition, anthropogenic sources associated with agricultural, industrial and urban land use are becoming increasingly prevalent (Korostynska et al., 2012). This includes the combustion of fossil fuels, which contributes substantially to atmospheric deposition (Boumans et al., 2004). In its bio-available form, nitrate is assimilated for growth and metabolism by riverine biota (e.g. algae, macrophytes and some bacteria) that form the basal components of aquatic food webs (Camargo et al., 2005). However, an excess of nitrate can lead to problems associated with eutrophication, such as decrease in light infiltration and dissolved oxygen concentration (Boesch et al., 2009; Bricker et al., 2008). This can negatively impact the health of aquatic biota such as invertebrates and fish (Camargo & Ward, 1995; Davidson et al., 2017; Moore & Bringolf, 2018). Understanding the dynamics of nitrate concentration in rivers, and the relationships nitrate has with other water-quality variables, is therefore of primary importance for the effective management of freshwater ecosystems.

Monitoring is central to understanding the links between water-quality variables and the health of freshwater ecosystems (Lessels & Bishop, 2013). Advances in the development of *in-situ* environmental sensors have led to their world-wide and long-term use in environmental monitoring (Martinez et al., 2004; Hart & Martinez, 2006; Jones et al., 2015). Yet, water-quality monitoring in rivers still relies primarily on the time-consuming and expensive manual collection of samples, with the resultant data being sparse in space and time (Leigh, Alsibai, et al., 2019). Fortunately, relatively low-cost *in-situ* sensors and sampling methods are being developed that allow some properties such as water temperature, turbidity, oxygen concentration, salinity and conductivity to be semi-continuously sampled and then analysed statistically (Park et al., 2020; Rodriguez-Perez et al., 2020). These high-frequency data-sets provide unique opportunities to better understand water-quality dynamics.

The large data-sets generated by *in-situ* sensors also present new challenges when analysing, modelling and reporting water-quality data (Park et al., 2020; Cawley, 2018); for example in terms of quality assurance and control (Leigh, Alsibai, et al., 2019). The prohibitive cost of certain sensors (Ruddell et al., 2014), such as optical sensors used to estimate high-frequency nitrate ($NO_3^-$) concentration, mean that they can only be deployed at a small number of sites and/or for limited periods of time. An additional challenge is to develop transferable models of water-quality dynamics and the links among water-quality variables for disparate river systems, especially for properties of interest like nitrate (Leigh, Kandanaarachchi, et al., 2019), especially when sensors are sparsely deployed. Ubiquity in water-quality relationships across climate zones, seasons and watersheds have thus far remained difficult to detect and model due to the many different processes that can be responsible for the resultant dynamics (Chanat et al., 2002).

While several other studies have explored models for nitrate concentration in rivers, they have been developed for different purposes, many for prediction rather than to understand underlying relationships among water-quality variables, and with varying degrees of success. For example, Diamantopoulou et al. (2005) used Artificial Neural Networks to predict monthly values of nitrate from multiple measures, including concentrations of other nutrients measured using traditional sampling and laboratory analyses. More recently, Leigh, Kandanaarachchi, et al. (2019) predicted nitrate from non-nutrient-based water-quality data collected from high-frequency, *in-situ* sensors using generalised-linear mixed-effects models (GLMMs) with a continuous first-order auto-regressive correlation (AR(1)) structure to account for temporal auto-correlation. However, GLMMs detect linear relationships and, as pointed out by Harrison et al. (2020), the relationships between nitrate and other water-quality variables tend to be nonlinear. Harrison et al. (2020) also investigated relationships between nitrate and other water-quality variables using high-frequency sensor data with the aim of prediction using Random Forests Regression (RFR) models, which can handle non-linear interactions among variables. However, predictions could not be extrapolated beyond the ranges of the input data and the resultant structure of the models was relatively opaque (Harrison et al., 2020). Links between discharge (also known as flow) and nitrate have also been investigated. However this relationship is complex, not least due to nutrient spiralling in streams (Ensign & Doyle, 2006), with some studies showing poor ability to model and explain variation in nitrate (Bowes et al. (2015), Wade et al. (2012)) while others finding much stronger relationships (Dupas et al. (2016)).

Our goal is to explore opportunities and address some challenges associated with high-frequency *in-situ* monitoring data. More specifically, we identified key variables and developed an additive model structure, which we used to understand the complex relationships between nitrate concentration and other water-quality variables, rather than as an exercise in prediction, collected in disparate climatic regions and subject to different levels of anthropogenic impacts.

## 2 Materials and methods

### 2.1 NEON database and water-quality sensors

The National Ecological Observatory Network (NEON) database provides open data from sites across the United States of America (USA). All NEON sites are equipped with high-frequency sensors and follow standardised configuration, calibration and preventive maintenance procedures (Vance et al., 2019; Willingham et al., 2019), with *in-situ* measurements and sample analyses following protocols outlined in Cawley (2016). Nitrate is measured in $\mu$M using a 10 mm path length SUNA V2 UV light spectrum sensor. The SUNA V2 collects data reported as a mean value from 20 measurements made during a sampling burst every 15 minutes (Table 1). Other sensors collect specific conductance ($\mu$S/cm), dissolved oxygen (mg/L), temperature (°C) and turbidity (Formazin Nephelometric Units,

FNU) data as one-minute instantaneous measurements. Water elevation (i.e. meters above sea level) data are also recorded as five-minute instantaneous measurements (Table 1).

Table 1. Details on NEON sensors, variables collected, units of measurement, associated data-collection intervals, and the NEON data product number for data used in this study.

| Sensor | Water-quality variable | Unit | Raw sensor collection interval | Published interval | NEON data product number |
|---|---|---|---|---|---|
| SUNA v2 | nitrate | $\mu$mol/L NO3-N | average of 20 bursts every 15 minutes | 15 minute | DP1.20033.001 |
| Level TROLL 500 | water level | meters above sea level | 1 minute | 5 minute | DP1.20016.001 |
| YSI EXO Optical Dissolved Oxygen | dissolved oxygen | mg/L DO | 1 minute | 1 minute | DP1.20288.001 |
| YSI EXO Turbidity | turbidity | FNU | 1 minute | 1 minute | DP1.20288.001 |
| YSI EXO Conductivity and Temperature | specific conductance | $\mu$ S/cm | 1 minute | 1 minute | DP1.20288.001 |
| Platinum Resistance Thermometer | temperature | $^\circ C$ | 1 minute | 1 minute | DP1.20053.001 |

### 2.2 Study sites and time-series data

We extracted time-series of nitrate (National Ecological Observatory Network, 2021b) and other surface water-quality variables (National Ecological Observatory Network, 2021d,c,a) from three different sites within the NEON database on 29 January 2021 (see Table 1): the Arikaree River in Colorado, Caribou-Poker Creeks Research Watershed in Alaska, and Lewis Run in Virginia, which we will refer to as Arikaree, Caribou and Lewis Run, respectively throughout (Table 2). Although nitrate measurements were collected more frequently than other water-quality variables (see Table 1), all sensors took measurements each time nitrate was sampled. Water-quality measurements from other time points were therefore discarded so that we only used data measured every 15 minutes (i.e. at each nitrate-measurement time point) in the analyses. Prior to analyses, we also removed data labelled as anomalous (e.g. due to known sensor calibration problems) during the rigorous NEON quality assurance and quality control procedure (Cawley, 2018). This, along with periods of missing data resulting from flow intermittence or sensors being temporarily out of service, meant that the time series of data from each site differed. For example, the river at Caribou freezes from approximately October to May each year and NEON removes most sensors from the site to prevent damage or loss. Despite these gaps in the data, at least 50% of the time series we examined from any one site overlapped with that of the other sites. Table 2 provides details about the selected data time series for each site, along with the actual time-period that could be used for modelling due to missing data. The time series of data are also available in the supplementary materials.

**Table 2.** Details of NEON sites used in this study.

|  | **Arikaree** | **Caribou** | **Lewis Run** |
|---|---|---|---|
| Watershed (km$^2$) | 2,875 | 106 | 11.9 |
| Manager | The Nature Conservancy | Bonanza Creek Long-Term Ecological Research Program and University of Alaska Fairbanks | Casey Trees (nonprofit organisation) |
| Climate zone | Semi-arid | Subarctic | Temperate |
| Land use / land cover | Grasslands, agriculture | Subarctic taiga, discontinuous permafrost | Fields, pastures, woodlands and small ponds |
| Flow persistence | Intermittent (dry in summer) | Perennial by ice covered in winter | Perennial |
| Mean annual precipitation (mm) | 450 | 262 | 976 |
| Urbanisation present in watershed? | No | No | Yes |
| Period of data analysed | January 2018 to December 2019 | January 2018 to December 2019 | January 2018 to December 2019 |
| Period of data modelled | September 2018 to December 2019 | June 2018 to October 2019 | January 2018 to December 2019 |

### 2.3 Statistical analyses

NEON publishes many environmental data products. The "water-quality" data product (National Ecological Observatory Network, 2021d) includes high-frequency pH, dissolved oxygen, oxygen saturation, turbidity, specific conductance, conductivity, chlorophyll-$a$, and fluorescent dissolved organic matter (fDOM) data streams. The "temperature (PRT) in surface water" product (National Ecological Observatory Network, 2021c) contains high-accuracy temperature data and the "surface water elevation" product (National Ecological Observatory Network, 2021a) includes elevation data derived from pressure transducers and site-surveyed elevations. Among the water-quality variables, many exhibit strong correlation with each other. We investigated multicollinearity in order to select only those variables that were independent or weakly correlated with each other. Multicollinearity between covariates can influence parameter estimates and inflate variances, leading to improper inference from fitted models (Kroll & Song, 2013), especially when the sample size is small. Although the data sets in this study were large ($n = 70\,080$ at Caribou, Arikaree and Lewis Run), we checked for multicollinearity using the variance inflation factor (VIF) (Helsel & Hirsch, 1992) to identify and remove any covariates that were strongly multi-collinear (VIF $\geq 6$). on this rule, we decided to not use conductivity, fluorescent dissolved organic matter (fDOM) pH and oxygen saturation. Also, the chlorophyll-$a$ data were not taken into account in this study due to their being an excessive number of anomalies. Therefore, we considered the following variables for further study: elevation of surface water, temperature, specific conductance and turbidity.

For the purposes of statistical analyses, we considered nitrate as the response (i.e. dependent variable) and the other water-quality variables (see Table 1) as covariates (i.e. explanatory variables or predictors). We also added a covariate for time in order to explore and account for the potential effect of season on nitrate concentration. Visual examination of the distributions of the response and covariates indicated that turbidity had a strongly right-skewed distribution and was therefore log-transformed (i.e. log (turbidity + 1)) prior to analysis.

Generalised additive mixed models (GAMMs) (Hastie & Tibshirani, 1990) were built to link nitrate concentration with covariates from each site individually as described by the equation:

$$Y_i = \beta_0 + \sum_{k=1}^{m} s_k(z_{ki}) + \eta_i \qquad (1)$$

where $z_{ki}$ are covariates measured at the $i$th sample ($i = 1, \ldots, n$). Here, $\beta_0$ is an intercept and $\eta_i$ is the auto-regressive moving-average (ARMA) ($\rho,\phi$) error, $\eta_i = \varepsilon_i + \sum_{j=1}^{p} \rho_j \eta_{i-j} + \sum_{l=1}^{q} \phi_l \varepsilon_{i-l}$ with $\rho_j$ the autocorrelation parameters and $\phi_l$ the moving average parameters, and $\varepsilon_i \overset{i.i.d}{\sim} \mathcal{N}(0, \sigma^2)$ is Gaussian white noise. The associated smooth function $s_k(\cdot)$ of each water-quality variable $z_k$ was defined using thin plate spline regression (Wood, 2017).

This model defined in (1) is estimated using a two-step modelling framework:

1. Step one: A generalised additive model (GAM) is used to model potential non-linear links between nitrate concentration and covariates. A stepwise variable-selection procedure was implemented and the 'best' GAM (variables and penalisation of smooth splines) for each site was identified based on the Akaike Information Criterion (AIC) (Akaike, 1979). To avoid over-fitting of the GAM model, the maximal value of degree of freedom of the smooth terms were fixed at 6.
2. Step two: An autoregressive–moving-average model was fitted to the GAM residuals to take into account the structural dependence of the time series data (GAMM)
The best ARMA regression on the GAM residuals were identified, based on the Akaike Information Criterion, to account for temporal autocorrelation in the time series data.

We then assessed the statistical significance and importance of each covariate in the best models to better understand their effects on nitrate. Variable importance can be estimated easily with linear models using the partial $R^2$ (i.e. the proportion of variation explained by a covariate in a model), but this approach cannot be used in a GAMM. Therefore, we compared the deviance explained by the best GAMM and the same GAMM with each water-quality covariate iteratively removed. We choose to present deviance explained in this paper to compare models and variable importance at each site because deviance explained is a percentage (restricting it between 0 and 100) making it easy to interpret. Finally, to compare performance of GAM versus GAMM in each site, we assessed the approximate Akaike Information Criterion (aAIC) given by the formula:

$$aAIC = n \log(\hat{\sigma}^2) + 2k \qquad (2)$$

(Burnham & Anderson, 2002) where $n$ is the length of the time series, $\hat{\sigma}^2$ is the variance of the model residuals and $k$ is the total number of degree of freedom in the model. We chose this approach because it is computationally efficient compared to other cross-validation methods (e.g. with multiple training/test set splits) and has reliable convergence properties asymptotically equivalent to one-step time series cross-validation. The smaller the aAIC, the better the model performance.

All analyses were undertaken in R statistical software (R Core Team 2020) using the `car` (Fox et al., 2020), `gam` (Hastie, 2020), `mgcv` (Wood, 2017), and `forecast` (Hyndman et al., 2021) packages. The R script used to implement the analyses is provided in supplementary materials and in the GitHub repository: `https://github.com/Claire-K/nitrate_links/blob/main/script_publi_arik.R`.

## 3 Results

### 3.1 Water-quality characteristics within and among sites

Each site had distinct water-quality characteristics (Figures 1 and 2). Lewis Run had a nitrate concentration far higher than Caribou and Arikaree sites (median = 5.5, 28.4 and 192 $\mu$M at Arikaree, Caribou and Lewis Run, respectively). For specific conductance, the Caribou site differed from the two other sites having, lower values (median = 531.6, 77.15 and 577.8 $\mu$S/cm at Arikaree, Caribou and Lewis Run, respectively). Dissolved oxygen concentration also differed among the three sites (median = 7.3, 12.34 and 9.55 mg/L at Arikaree, Caribou and Lewis Run, respectively) water-quality variables exhibited more variability at Arikaree and Lewis Run than at Caribou (Figure 1). Temperature ranges were narrowest at Caribou (0 to 13 °C), wider at at Lewis Run (1 to 22 °C), and widest Arikaree (0 to 34 °C). As noted above, turbidity was strongly right-skewed in distribution at all three sites, with the mean always greater in value than the third quartile (mean = 95.45 FNU, Q3 = 9.98 FNU at Arikaree; mean = 3.41 FNU, Q3 = 3.15 FNU at Caribou; mean = 23.85 FNU, Q3 = 23.93 FNU at Lewis Run). Finally, surface-water elevation was very different among sites, and despite the small ranges between minimum and maximum elevations, exhibited some temporal variability at Caribou (230.0 - 230.8 m), Arikaree (1179 - 1180 m) and Lewis Run (125.9 - 126.7) during the study period.

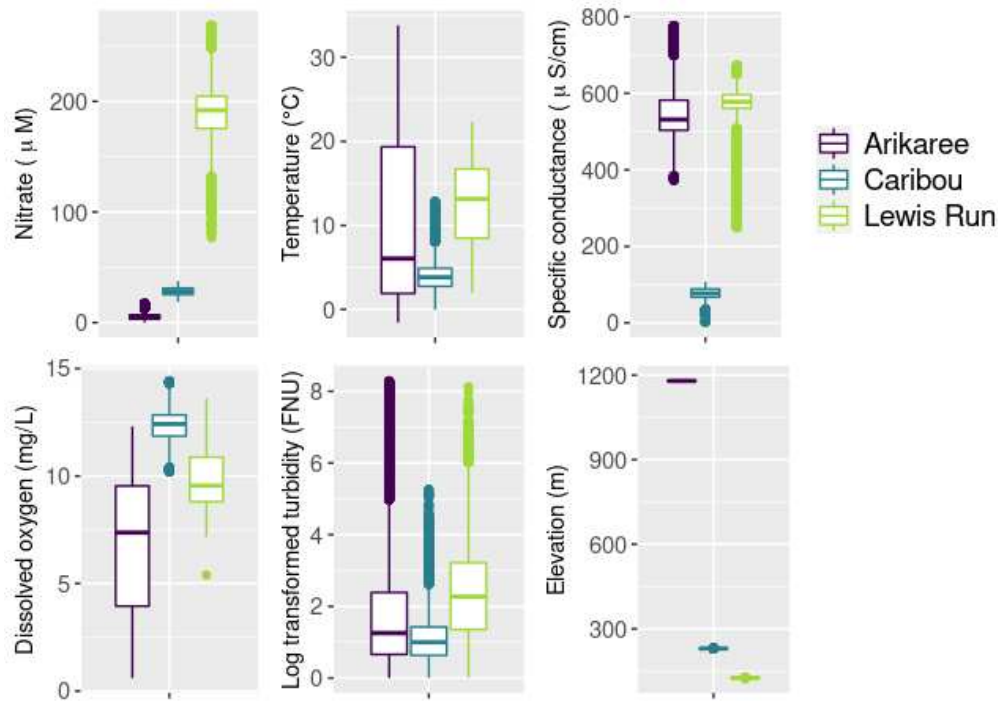

**Figure 1.** Box-plots of water-quality data for Arikaree, Caribou and Lewis Run. Bold lines within boxes represent medians and lower and upper edges of boxes represent the interquartile range (IQR), with whiskers extending to 1.5 times the IQR. Closed circles represent data with values beyond the whiskers.

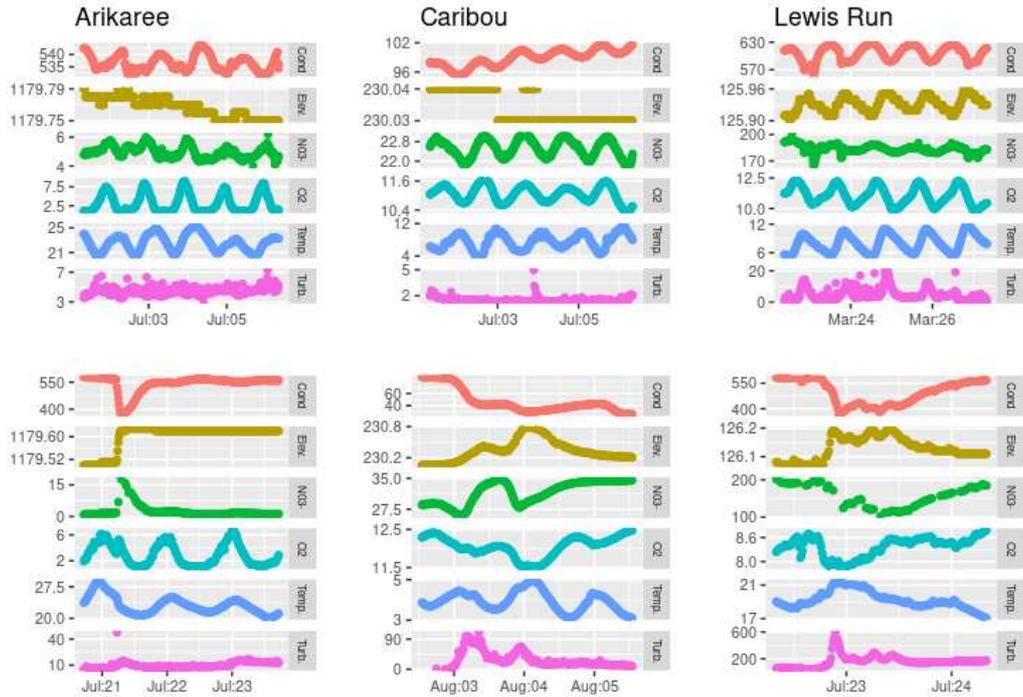

**Figure 2.** The first three plots are examples of diel fluctuations and other trends in water-quality at the study sides, visualised over 5-day windows of representative data, each starting and ending at 10 pm GTM. The three bottom plots are examples of flow events, which are indicated by sudden rises in the water elevation. "Cond.", "Elev.", "$NO3^-$", "$O_2$", "Temp" and "Turb." respectively correspond to specific conductance, elevation, nitrate concentration, dissolved oxygen concentration, temperature and turbidity, which is on the logarithmic scale. Units of measurement are as per Table 1. Note that the scales for the y-axes differ among sites.

Diel fluctuations in water-quality occurred at all three sites despite differences in the distributions of water-quality data among them (top plots of Figure 2). Data from NEON are given in GTM, Arikaree is GTM-7, Caribou is GTM-8 and Lewis Run is GTM-5. Local time at the beginning at the three top plots are 3PM at Arikaree, 2PM at Caribou, and 5 PM at Lewis Run. At Caribou, nitrate, dissolved oxygen, specific conductance and turbidity increased while temperature decreased during the afternoon and the night. At Arikaree, the diel patterns exhibited comparatively more variation than at Caribou, and both nitrate and dissolved oxygen fluctuated in the opposite direction (i.e. decreasing at night and increasing during the day). Lewis Run exhibited the same diel patterns as Caribou and Arikaree, with a clear alternation between day and night in turbidity.

When a flow event (lower plots in Figure 2) occurred at Arikaree site (i.e. when the elevation level suddenly rose), nitrate concentration, oxygen concentration and turbidity increased, while specific conductance and temperature decreased. Conversely, at Lewis Run, a sudden increase in water level coincided a decrease in nitrate concentration. At Caribou, the rise in water level was accompanied by an increase in turbidity, but the relationships between water level and the other water-quality covariates were more complex.

### 3.2 GAMs

GAMs regressions were used to understand the links between nitrate concentration and each covariate (Figure 3). The smooth regressions between nitrate and the covariates revealed differences among sites and covariates. For example, the time smooth regression (i.e. the expected change in nitrate concentration) peaked around April-May (spring) and September-October (fall) at Arikaree, and in winter at Lewis Run, whereas it exhibited an increasing trend during fall at Caribou. The smooth regression of temperature (Figure 3) demonstrated a light negative effect on the expected change in nitrate concentration at Caribou and Arikaree. The reverse occurred at Lewis Run, with the expected change in nitrate concentration increasing with temperature.

For dissolved oxygen, the smooth regression indicated a minimal effect on expected change in nitrate concentration at Arikaree (Figure 3). This was also the case for Caribou up until a dissolved oxygen concentration of 13.5 mg/L when the expected change in nitrate concentration increased sharply; however, this increase was due to a small number of high-concentration dissolved oxygen measurements only. The relationship between expected change in nitrate concentration and dissolved oxygen at Lewis Run was also weak.

The specific conductance smooth regression had a very different effect on the expected change in nitrate concentration at all three sites. At Caribou, there was a strong, negative effect on expected change in nitrate concentration between until 55 $\mu$S/cm followed by a strong, positive effect between about 55 to 65 $\mu$S/cm when a more gentle, negative effect returned. Ranges of specific conductance at Arikaree and Lewis Run were similar, but smooth regressions were different for small values. At Arikaree, the expected change on nitrate concentration increased until $500\mu$S/cm whereas it decreased at Lewis Run. In the two sites, when the specific conductance was higher than 550 $\mu$S/cm, nitrate concentration was expected to increase.

Confidence intervals for the smoothed regressions of log-transformed turbidity on the expected change in nitrate concentration tended to be wide in all three models compared with those of other covariates (Figure 3). The effect of turbidity, although weak, tended overall to be negative at Caribou and Arikaree, whereas at Lewis Run the effect tended to be more variable.

The relationship between the smooth regression for water elevation and expected change in nitrate concentration was also different among sites. At Caribou, the expected change in nitrate concentration tended to increase with elevation while the reverse was observed at Lewis Run. At Arikaree, confidence intervals were wide at the start of the time series and cannot be interpreted, leading the nitrate concentration tended to increased with water elevation, just as at Caribou.

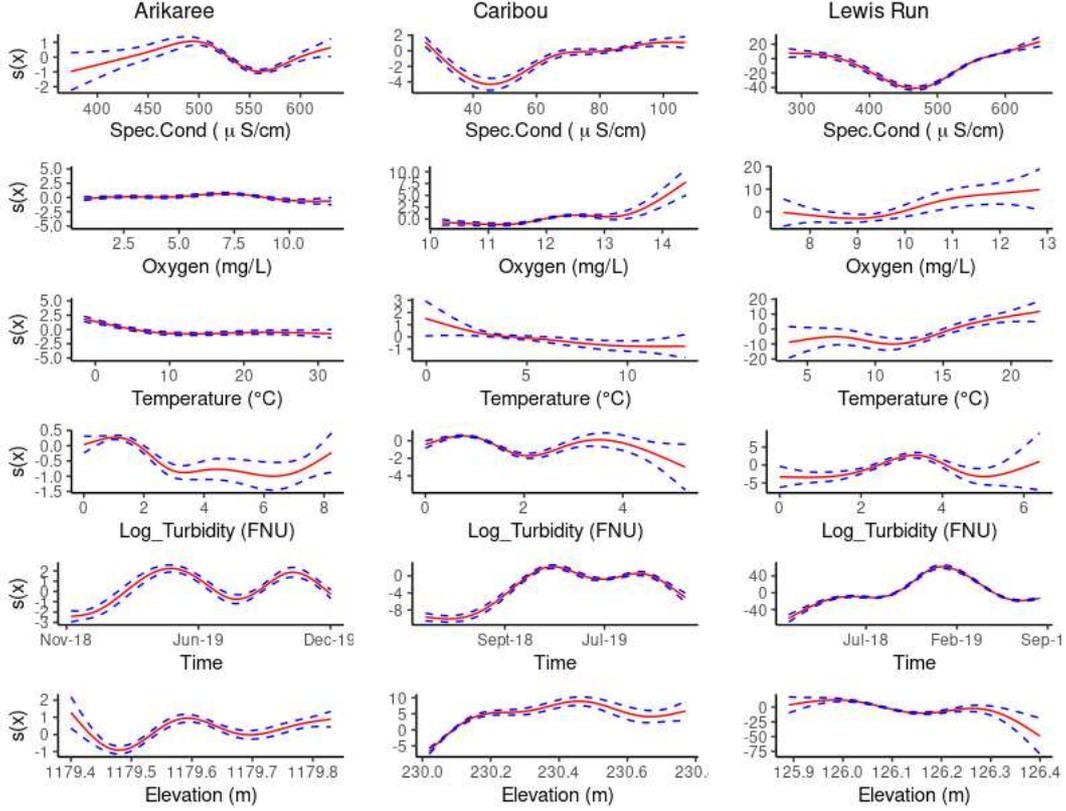

**Figure 3.** Smooth regressions (solid lines) of the expected change in nitrate concentration (y-axes) when values taken by covariates (x-axis) increased, by site. More specifically, each plot represents the regressive spline $s_k(z_{ki})$ (Equation (1)) of one covariate at one site, i.e. the expected change in nitrate concentration $Y_i$ for a one unit increase in the covariate $z_{ki}$. Dashed lines show the standard error estimates. Missing values (gaps in the time series data) are not shown.

### 3.3 GAMMs

The two-step modeling framework used in this study allowed us to differentiate the amount of deviance explained by covariates (GAM step) and the amount of deviance explained by the auto-regressive model (ARMA step). In the first step, we included all variables as potential covariates in the GAMs because the VIF results indicated that multicollinearity among the covariates (the water-quality and time variables) was not a significant concern (VIFs < 6). The best GAMs, as selected by the stepwise variable-selection procedure based on the AIC, achieved a deviance explained of 74%, 92% and 83% respectively for Arikaree, Caribou and Lewis Run (i.e. the final GAMMs) to 99% for all three sites. The auto-regressive ARMA functions then fit to the residuals of the best GAMs were ARMA(2,3), AR(3,2) and ARMA(2,1), respectively, for Arikaree, Caribou and Lewis Run, which increased the total deviance explained by the models.

**Table 3.** Model performance for all three sites, as based on the approximated Akaike Information Criterion (aAIC).

| Site | Model | aAIC |
|---|---|---|
| Arikaree | GAM | 6671 |
|  | GAMM | -44696 |
| Caribou | GAM | 2220 |
|  | GAMM | -17280 |
| Lewis Run | GAM | 61408 |
|  | GAMM | 31190 |

Performance of models built for each sites were compared using the aAIC (Equation 2). For the three sites, GAMMs performed far better than GAMs, with the GAMMs aAIC. However, the GAMs all explained a large proportion of variation in nitrate with the same combination of covariates, regardless of site. These included smooth terms for specific conductance, dissolved oxygen, temperature, turbidity (log-transformed), time and water elevation, which were all statistically significant ($p < 0.001$) (Figure 4). However, the importance of covariates in the models fit to nitrate data differed among sites. Overall, the relatively low importance of all covariates at Arikaree compared to that at Caribou and Lewis Run (Figure 4) was congruent with the amount of deviance explained by the GAM portion of the Arikaree model (74%), it being the lowest among sites. The most important variable in the GAM for Arikaree was water temperature, but it explained only 7% of the deviance in the final GAMM. However, the auto-regressive portion of the GAMM for Arikaree was particularly important, explaining 25% of the deviance. At Lewis Run, all the water-quality variables were important in explaining nitrate. The variables with the greatest importance were specific conductance and time (almost 15% each). The importance of all other water-quality covariates were between 10% and 12% of the nitrate concentration deviance. These higher values of variable importance at Lewis Run corresponded with the larger proportion of deviance explained (83%) by the GAM part of the GAMM for this site. At Caribou, the auto-regressive part of the GAMM was not very important in explaining nitrate concentration (only 7%) while the amount of deviance in nitrate concentration explained by the GAM portion of the model was high (92%). Time and water surface elevation were the most important variables (12% each) in the Caribou model.

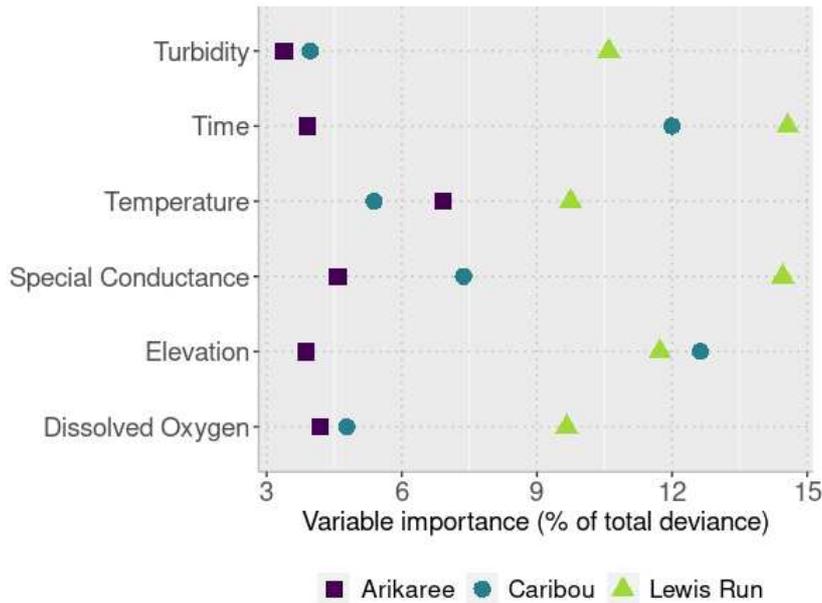

**Figure 4.** Variable importance, as the percentage of the total GAMM model deviance, for statistically significant covariates (p < 0.01), by study site. Total deviance explained by each GAMM was 99% for all three sites

## 4 Discussion

Our study has demonstrated that GAMMs provide a suitable and useful method to model and understand the nonlinear relationships between nitrate and other water-quality variables, with an ability to explain 99% of the variation in nitrate concentration. Leigh, Kandanaarachchi, et al. (2019) achieved < 22% of deviance explained with GLMMs using a continuous first-order auto-regressive correlation (AR(1)). However, GLMMs detect linear relationships and, as pointed out by Harrison et al. (2020) and demonstrated by our study, the relationships between nitrate and other water-quality variables tend to be nonlinear. This, along with the complexity of nitrate dynamics in rivers (Bowes et al., 2015), may help to explain why the GAMMs we fit explained a high proportion of variation in nitrate (99%) compared with the GLMMs (< 22%; Leigh, Kandanaarachchi, et al. (2019)). Random Forests Regression (RFR) models can handle nonlinear interactions among variables and have been shown to explain 89% of nitrate concentration (Harrison et al., 2020). However, Harrison et al. (2020) built their RFR models for the purpose of prediction and indicate that the resultant structure of the models was relatively opaque. Thus, we suggest GAMM regression in preference to other models like GLMMs and RFR when the primary aim is to explore and better understand the links between nitrate and other water-quality variables. The two-step approach we used in this study is also highly relevant because, in addition to improving the understanding of links between nitrate and other water-quality variables, it enables the importance of temporal auto-correlation in nitrate measurements to be accounted for and assessed.

The use of GAMMs with big data has nevertheless raised computational challenges. The first challenge was the presence of missing data and technical anomalies in the time series. Within the two-year period of data analysed for this study, there were extended periods of such data that could not be incorporated into the models, which constrained the lengths of the time series modelled as a result (see supplementary materials). A second challenge was building the GAMMs themselves. The most commonly used function in the R `mgcv` package to fit GAMMs is 'gamm', which enables an ARMA process to be fit to GAM residuals. However, as explained in the gammmgcv vignette (*gammmgcv: Generalized Additive Mixed Models*, n.d.), 'gamm' is typically much slower to run than the 'gam' function, and the amount of memory required by R to run 'gamm' with large data sets may need to be increased substantially. Using the large data sets in this study often resulted in function failure, which required us to individually code the two-step GAMMs. This meant that temporal auto-correlation was not accounted for in the smoothing-parameter selection step, such that any auto-correlation could potentially be confused with trend resulting in under-estimation of the smoothing parameters and bias during inference. However, a two-step approach, such as that we used to secondarily account for auto-correlation, while typically not as efficient as estimating auto-correlation and smoothing parameters at the same time, is often more robust (Casella & Berger, 2002). A final computational challenge in this study was to calculate variable importance for the GAMs. Several solutions are available to calculate variable importance for linear models (see, for example the `vimp` package (Williamson et al., 2020)), but, to our knowledge, no solutions are available to easily calculate variable importance for non-linear models. As a result, we needed to determine variable importance using the relatively time-consuming method of iteratively comparing models with and without each covariate.

In terms of describing and better understanding nitrate dynamics and the relationship between nitrate concentration and other water-quality variables, the models we developed indicated similarities and differences among sites. Firstly, sites differed in their nitrate dynamics, likely relating to the distinct environmental conditions of the regions in which each sites was located. Rivers in regions with discontinuous permafrost, like Caribou, tend to export nitrogen (including in dissolved form, i.e. nitrate) rather than retain it, which is more typical of rivers in temperate regions like Arikaree and Lewis Run (Jones Jr. et al., 2005). At Caribou, water elevation was the among most important variable explaining variation in nitrate concentration, along with time. Water elevation was positively correlated with nitrate (when elevation increased so did nitrate concentration), it is likely that the increased stream level is from rain storms, which are serving to flush nitrate from shallow flow-paths through the watershed. Another hypothesis is that the positive link between water elevation and nitrate concentration could be due to induced daily cycles of evapotranspiration (Yashi et al., 1990) in water surface elevation resulting in nitrate dissolution in water. It is also possible that algae and other photosynthetic microorganisms active during the day were depleting nitrate. At Arikaree, water elevation was not important in explaining nitrate fluctuations, with a wide confidence interval around the smooth regression. This suggests that daily fluctuations in nitrate at Arikaree were not due to water elevation changes and/or evapotranspiration. The most likely hypothesis is that daily fluctuations in nitrate were due to the activity of algae and other photosynthetic microorganisms that use nitrate as an essential element for photosynthesis.

On the whole, water-quality variables at Caribou tended to be reflective of the relative stability of the subarctic Alaskan ecosystem in which the site is located, being in a reserve upstream of urbanisation, with no known anthropogenic pollution present in chemical or physical form, including in the nitrate delivered into the system via atmospheric deposition (which has been monitored at the site since 19993 as part of the United States National Atmospheric Deposition Program. In contrast, multiple peaks and/or troughs in nitrate, turbidity and conductance occurred at Lewis Run, which could be associated with its proximity to an urban area (Washingtom, D.C.), a water treatment plant and location within a predominantly agricultural watershed, which could not only increase the overall nitrate concentration at this site but also the temporal variability in water-quality (Blann et al., 2009). In fact, the highest concentration of nitrate among the sites was observed at Lewis Run. The relatively high temporal variability in nitrate, turbidity and conductance at Lewis Run may also be affected by climate, being in a temperate zone with large thermal and rainfall amplitudes occurring between summer and winter. water-quality at Arikaree also fluctuated substantially, and like Lewis Run, is located in an agricultural region, albeit within a semi-arid climate zone.

This work provides valuable insight on the links between nitrate and other water-quality variables in river systems, using approximately two years of data from each of three sites. While any analysis will only capture the range of variation contained within the input data itself, our choice of sites being distinct in their environmental and flow-regime characteristics, and the high-frequency nature of the sensor-based water-quality data that we analysed, means that inference can be drawn across a range of conditions. Continual analysis of data collected over future years, may, however, reveal new patterns and trends that allow for a more in-depth understanding of the relationships among variables. On the other hand the computational effort needed to create models from several years of data is large, and the availability of time series from high-frequency sensors that are absent from long or multiple sequences of data is limited. Nevertheless, testing the predictive ability of the models developed herein is likely to provide further insight on the links among water-quality variables while also enabling the prediction, for example, of missing or anomalous data in sensor-based time series.

Despite the among-site differences in nitrate concentration and other water-quality variables, the final GAMMs for each site included the same set of water-quality covariates. This demonstrates that these water-quality variables are consistently important for understanding variation in nitrate in rivers, even in watersheds with different types of land use and in different climate zones. The transferability of models, for example between different sites, remains a challenging obstacle in environmental and ecological modelling, as does the evaluation of their transferability (Yates et al., 2018). However, our results suggested that a single model was not appropriate for the sites we examined, given the site-specific differences in relationship between nitrate and the other water-quality variables. Rather, we were able to identify a transferable modelling framework and a set of common of covariates that could together be used to explain nitrate concentration across disparate sites. Although GAMMs in such a framework must be tailored to data from individual sites, future research may reveal that models fit to data from sites with more similar land use, climate conditions and flow regimes are more transferable. With the implementation of automated sensing across several sites, watersheds and potentially regions, producing transferable models will become increasingly sought after in order to better understand water-quality relationships and dynamics, and to support water-resource management (Leigh, Alsibai, et al., 2019).

The findings of this study are highly relevant for scientists and managers responsible for *in-situ* monitoring in rivers. As mentioned above, gaps in *in-situ* sensor data are common and the methods demonstrated here can be used to impute missing data and provide a more holistic description of nitrate dynamics. This is particularly important when financial resources are limited and decisions must be made about which sensors to buy and which water-quality variables to measure. In addition, this work provides a basis for future studies focused on the prediction of other critically important, water-quality variables that cannot be measured using *in-situ* sensors or when the sensors themselves are cost prohibitive.

**Acknowledgments**

Funding was provided by the Energy Environment Solutions (E2S-UPPA) consortium thought an international research chair. This study was part of a project funded by an Australian Research Council (ARC) Linkage grant (LP180101151) "Revolutionising high resolution water-quality monitoring in the information age", and we extend our thanks to all the team members involved in the broader project. Data is available through (National Ecological Observatory Network, 2021b) (nitrate concentration) and (National Ecological Observatory Network, 2021d,c,a) (other surface water-quality variables). The National Ecological Observatory Network is a program sponsored by the National Science Foundation and operated under cooperative agreement by Battelle Memorial Institute. This material is based in part upon work supported by the National Science Foundation through the NEON Program.

**References**

Akaike, H. (1979, 08). A Bayesian extension of the minimum AIC procedure of autoregressive model fitting. *Biometrika*, *66*(2), 237-242. doi: 10.1093/biomet/66.2.237

Blann, K. L., Anderson, J. L., Sands, G. R., & Vondracek, B. (2009). Effects of agricultural drainage on aquatic ecosystems: a review. *Critical reviews in environmental science and technology*, *39*(11), 909–1001. doi: 10.1080/10643380801977966

Boesch, D. F., Boynton, W. R., Crowder, L. B., Diaz, R. J., Howarth, R. W., Mee, L. D., ... others (2009). Nutrient enrichment drives Gulf of Mexico hypoxia. *Eos, Transactions American Geophysical Union*(14), 117–118. doi: 10.1029/2009EO140001

Boumans, L., Fraters, D., & Van Drecht, G. (2004). Nitrate leaching by atmospheric N deposition to upper groundwater in the sandy regions of the Netherlands in 1990. *Environmental monitoring and assessment*, *93*(1-3), 1–15. doi: 10.1023/B:EMAS.0000016788.24386.be

Bowes, M., Jarvie, H., Halliday, S., Skeffington, R., Wade, A., Loewenthal, M., ... Palmer-Felgate, E. (2015). Characterising phosphorus and nitrate inputs to a rural river using high-frequency concentration–flow relationships. *Science of The Total Environment*, *511*, 608-620. doi: 10.1016/j.scitotenv.2014.12.086

Bricker, S. B., Longstaff, B., Dennison, W., Jones, A., Boicourt, K., Wicks, C., & Woerner, J. (2008). Effects of nutrient enrichment in the nation's estuaries: a decade of change. *Harmful Algae*, *8*(1), 21–32. doi: 10.1016/j.hal.2008.08.028

Burnham, K. P., & Anderson, D. R. (2002). *Model selection and multimodel inference: a practical information-theoretic approach.* Springer-Verlag.

Camargo, J. A., Alonso, A., & Salamanca, A. (2005). Nitrate toxicity to aquatic animals: a review with new data for freshwater invertebrates. *Chemosphere*, *58*(9), 1255–1267. doi: 10.1016/j.chemosphere.2004.10.044

Camargo, J. A., & Ward, J. (1995). Nitrate (NO3-N) toxicity to aquatic life: A proposal of safe concentrations for two species of nearctic freshwater invertebrates. *Chemosphere*, *31*(5), 3211–3216. doi: 10.1016/0045-6535(95)00182-8

Casella, G., & Berger, R. L. (2002). *Statistical inference.* Pacific Grove, CA: Duxbury.

Cawley, K. M. (2016). *Neon aquatic sampling strategy NEON.DOC001152vA.* Boulder, CO, USA: National Ecological Observatory Network. Retrieved from http://data.neonscience.org

Cawley, K. M. (2018). *Neon algorithm theoretical basis document.* Boulder, CO, USA: National Ecological Observatory Network.

Chanat, J. G., Rice, K. C., & Hornberger, G. M. (2002). Consistency of patterns in concentration-discharge plots. *Water Resources Research*, *38*(8), 22–1. doi: 10.1029/2001WR000971

Davidson, J., Good, C., Williams, C., & Summerfelt, S. T. (2017). Evaluating the chronic effects of nitrate on the health and performance of post-smolt Atlantic salmon Salmo salar in freshwater recirculation aquaculture systems. *Aquacultural Engineering*, *79*, 1 - 8. doi: 10.1016/j.aquaeng.2017.08.003

Diamantopoulou, M. J., Papamichail, D. M., & Antonopoulos, V. Z. (2005). The use of a neural network technique for the prediction of water quality parameters. *Operational Research*, *5*(1), 115–125. doi: 10.1007/BF02944165

Dupas, R., Jomaa, S., Musolff, A., Borchardt, D., & Rode, M. (2016). Disentangling the influence of hydroclimatic patterns and agricultural management on river nitrate dynamics from sub-hourly to decadal time scales. *Science of the Total Environment*, *571*, 791–800. doi: 10.1016/j.scitotenv.2016.07.053

Ensign, S. H., & Doyle, M. W. (2006). Nutrient spiraling in streams and river networks.

*Journal of Geophysical Research: Biogeosciences*, *111*(G4). doi: 10.1029/2005JG000114

Fox, J., Weisberg, S., Adler, D., Bates, D., Baud-Bovy, G., Bolker, B., ... R-Core (2020). *car: Companion to applied regression.* Retrieved from https://CRAN.R-project.org/package=car (R package version 3.0-10)

*gammmgcv: Generalized additive mixed models.* (n.d.). https://stat.ethz.ch/R-manual/R-patched/library/mgcv/html/gamm.html. (Accessed: 2021-03-08)

Harrison, J. W., Lucius, M. A., Farrell, J. L., Eichler, L. W., & Relyea, R. A. (2020). Prediction of stream nitrogen and phosphorus concentrations from high-frequency sensors using random forests regression. *Science of The Total Environment*, 143005. doi: 10.1016/j.scitotenv.2020.143005

Hart, J. K., & Martinez, K. (2006). Environmental sensor networks: A revolution in the earth system science? *Earth-Science Reviews*, *78*(3), 177 - 191. doi: 10.1016/j.earscirev.2006.05.001

Hastie, T. (2020). *gam: Generalized additive models.* Retrieved from https://CRAN.R-project.org/package=gam (R package version 1.20)

Hastie, T., & Tibshirani, R. (1990). *Generalized additive models.* CRC press.

Helsel, D. R., & Hirsch, R. M. (1992). *Statistical methods in water resources.* Elsevier.

Hyndman, R., Athanasopoulos, G., Bergmeir, C., Caceres, G., Chhay, L., O'Hara-Wild, M., ... Yasmeen, F. (2021). *forecast: Forecasting functions for time series and linear models.* Retrieved from https://CRAN.R-project.org/package=forecast (R package version 8.14)

Jones, A. S., Horsburgh, J. S., Reeder, S. L., Ramirez, M., & Caraballo, J. (2015). A data management and publication workflow for a large-scale, heterogeneous sensor network. *Environmental monitoring and assessment*, *187*(6), 348.

Jones Jr., J. B., Petrone, K. C., Finlay, J. C., Hinzman, L. D., & Bolton, W. R. (2005). Nitrogen loss from watersheds of interior Alaska underlain with discontinuous permafrost. *Geophysical Research Letters*, *32*(2). doi: 10.1029/2004GL021734

Korostynska, O., Mason, A., & Al-Shamma'a, A. (2012). Monitoring of nitrates and phosphates in wastewater: Current technologies and further challenges. *International Journal on Smart Sensing & Intelligent Systems*, *5*(1). doi: 10.21307/ijssis-2017-475

Kroll, C. N., & Song, P. (2013). Impact of multicollinearity on small sample hydrologic regression models. *Water Resources Research*, *49*(6), 3756–3769. doi: 10.1002/wrcr.20315

Leigh, C., Alsibai, O., Hyndman, R. J., Kandanaarachchi, S., King, O. C., McGree, J. M., ... Peterson, E. E. (2019). A framework for automated anomaly detection in high frequency water-quality data from in situ sensors. *Science of the Total Environment*, *664*, 885 - 898. doi: 10.1016/j.scitotenv.2019.02.085

Leigh, C., Kandanaarachchi, S., McGree, J. M., Hyndman, R. J., Alsibai, O., Mengersen, K., & Peterson, E. E. (2019). Predicting sediment and nutrient concentrations from high-frequency water-quality data. *PLOS ONE*, *14*(8), e0215503. doi: 10.1371/journal.pone.0215503

Lessels, J., & Bishop, T. (2013). Estimating water quality using linear mixed models with stream discharge and turbidity. *Journal of Hydrology*, *498*, 13-22. doi: 10.1016/j.jhydrol.2013.06.006

Martinez, K., Hart, J. K., & Ong, R. (2004). Environmental sensor networks. *Computer*, *37*(8), 50-56. doi: 10.1109/MC.2004.91

Moore, A. P., & Bringolf, R. B. (2018). Effects of nitrate on freshwater mussel glochidia attachment and metamorphosis success to the juvenile stage. *Environmental pollution*, *242*, 807–813. doi: 10.1016/j.envpol.2018.07.047

National Ecological Observatory Network. (2021a). *Elevation of surface water (DP1.20016.001).* Boulder, CO, USA. Retrieved from https://data.neonscience.org/data-products/DP1.20016.001/RELEASE-2021 doi: 10.48443/QSER-8M94

National Ecological Observatory Network. (2021b). *Nitrate in surface water (DP1.20033.001).* Boulder, CO, USA. Retrieved from https://data.neonscience.org/

data-products/DP1.20033.001/RELEASE-2021 doi: 10.48443/924T-1K41

National Ecological Observatory Network. (2021c). *Temperature (PRT) in surface water (DP1.20053.001).* Boulder, CO, USA. Retrieved from https://data.neonscience.org/data-products/DP1.20053.001/RELEASE-2021 doi: 10.48443/NY19-PJ91

National Ecological Observatory Network. (2021d). *Water quality (DP1.20288.001).* Boulder, CO, USA. Retrieved from https://data.neonscience.org/data-products/DP1.20288.001/RELEASE-2021 doi: 10.48443/D8KW-5J62

Park, J., Kim, K. T., & Lee, W. H. (2020). Recent advances in information and communications technology (ICT) and sensor technology for monitoring water quality. *Water*, *12*(2), 510.

Rodriguez-Perez, J., Leigh, C., Liquet, B., Kermorvant, C., Peterson, E., Sous, D., & Mengersen, K. (2020). Detecting technical anomalies in high-frequency water-quality data using artificial neural networks. *Environmental Science & Technology*, *54*(21), 13719–13730. doi: 10.1021/acs.est.0c04069

Ruddell, B. L., Zaslavsky, I., Valentine, D., Beran, B., Piasecki, M., Fu, Q., & Kumar, P. (2014). Sustainable long term scientific data publication: Lessons learned from a prototype Observatory Information System for the Illinois River Basin. *Environmental Modelling & Software*, *54*, 73 - 87. doi: 10.1016/j.envsoft.2013.12.015

Vance, J., Nance, B., Monahan, D., Mahal, M., & Cavileer, M. (2019). *NEON preventive maintenance procedure: AIS surface water quality multisonde. NEON.DOC.001569vB.* Boulder, CO, USA: National Ecological Observatory Network. Retrieved from https://data.neonscience.org/data-products/DP1.20288.001/RELEASE-2021

Wade, A. J., Palmer-Felgate, E., Halliday, S. J., Skeffington, R. A., Loewenthal, M., Jarvie, H., ... others (2012). Hydrochemical processes in lowland rivers: insights from in situ, high-resolution monitoring. *Hydrology and Earth System Sciences*, *16*(11), 4323–4342. doi: 10.5194/hess-16-4323-2012

Williamson, B., Gilbert, P., Carone, M., & Simon, R. (2020). Nonparametric variable importance assessment using machine learning techniques. *Biometrics*. doi: 10.1111/biom.13392

Willingham, R., Cavileer, M., Csavina, J., & Monahan, D. (2019). *NEON preventive maintenance procedure: Submersible ultraviolet nitrate analyzer. NEON.DOC.002716vB.* Boulder, CO, USA: National Ecological Observatory Network. Retrieved from http://data.neonscience.org

Wood, S. N. (2017). *Generalized additive models: an introduction with R.* CRC press.

Yashi, D. K., Suzuki, K., & Nomura, M. (1990). Diurnal fluctuation in stream flow and in specific electric conductance during drought periods. *Journal of Hydrology*, *115*(1), 105 - 114. doi: 10.1016/0022-1694(90)90200-H

Yates, K. L., Bouchet, P. J., Caley, M. J., Mengersen, K., Randin, C. F., Parnell, S., ... Sequeira, A. M. (2018). Outstanding challenges in the transferability of ecological models. *Trends in Ecology & Evolution*, *33*(10), 790 - 802. doi: 10.1016/j.tree.2018.08.001